\documentclass[12pt]{article}
\usepackage{enumerate,amssymb}
\renewcommand{\a}{\alpha}
\renewcommand{\b}{\beta}
\newcommand{\g}{\gamma}           
\renewcommand{\d}{\delta}

\newcommand{\la}{\lambda}

\newcommand{\s}{\sigma}           
         
\newcommand{\f}{{\phi}}

\newcommand{\be}{\begin{equation}}
\newcommand{\ee}{\end{equation}}
\newcommand{\eqn}[1]{\label{#1}\end{equation}}

\newcommand{\bea}{\begin{eqnarray}}
\newcommand{\eea}{\end{eqnarray}}
\newcommand{\eqan}[1]{\label{#1}\end{eqnarray}}

\newcommand{\ba}{\begin{array}}
\newcommand{\ea}{\end{array}}

\newcommand{\nn}{\nonumber}

\begin{document}

\begin{center}
{\bf   Complete and Consistent Non-Minimal String Corrections to Supergravity}\\[14mm]

S. Bellucci\\

{\it INFN-Laboratori Nazionali di Frascati\\
Via E. Fermi 40, 00044 Frascati, Italy\\ mailto: bellucci@lnf.infn.it}\\[6mm]

D. O'Reilly\\

{\it Physics Department, The Graduate School and University
Center\\365 Fifth Avenue
New York, NY 10016-4309\\ mailto: doreilly@gc.cuny.edu }\\[6mm]

\end{center}
\vbox{\vspace{3mm}}

\begin{abstract}  We give a complete and consistent solution to string corrected (deformed),
 $D=10$, $N=1$ supergravity as the non-minimal low energy limit of string theory.
 We solve the Bianchi identities with suitable constraints to second order in the string slope parameter.
  In so doing we pave the way for continuing the study of the many applications of these results.
  We also modify, reaffirm and correct a previously given incomplete
  solution, and we introduce an important adjustment to the known first
  order results.

 \vbox{\vspace{1mm}}

\end{abstract}

\newpage

\section{Introduction}

This work is inspired by the celebration of the activity of Prof.
Buchbinder. One of us (S.B.) has been a steady collaborator of the
Tomsk group in the past decade, with 16 original papers published
in collaboration. Most of them involved Prof. Anton Galajinsky and
two were directly related to the work of Prof. Buchbinder. The
latter ones dealt with noncommutative field theory, however many
discussions and interactions with Prof. Buchbinder over the years,
concerned issues related to supersymmetry and supergravity
theories, an area where Prof. Buchbinder obtained some of his
numerous prestigious achievements. Hence it is quite suitable to
present in this dedicated volume an investigation that connects to
the stimulating and seemingly almost everlasting issue of
string-corrected ten-dimensional supergravity theories.

The route to finding a manifestly supersymmetric theory of D= 10,
N=1 supergravity at second order in the string slope parameter has
encountered many difficulties over the years. Some years ago a
solution to D=10, N=1 Supergravity as the low energy limit of
String Theory was given at first order in the string slope
parameter and was recently re-calculated \cite{1}. In a sense this
was a minimal solution. This approach was founded on what is
nowadays refered to, as the scenario of Gates and collaborators;
(see \cite{1}, \cite{2}, and references therein). Other varied
approaches are also pursued, however the power of this older
approach is now being vindicated. A partial second order solution
was recently given in \cite{3} and \cite{4}. It was incomplete and
therefore in doubt due do an unsatisfactory assumption in the
curvature sector, as well as a computational error. Here we
reaffirm that that solution is correct up to a curvature term, and in particular that
the proposed X tensor is valid.
 We then show that the results obtained can be used to solve the
curvature Bianchi identity,  equation (3). We achieve this by introducing
$R^{(1)}{}_{ a b \g}{}^{\d}$, and then imposing
 a condition on it which also modifies the old first order
results. The difficulties that prevented the complete closing of
the Bianchi identities at second order are overcome. We
present the full set of equations that consistently satisfy all
required Bianchi identities. As the work in itself is lengthy we leave
finding the equations of motion and other applications for another
letter. We also do not list results explicitly solved by Bianchi
identies. For this approach it is
required that we solve the Bianchi identities for D=10 N=1
supergravity in superspace at second order in the string slope parameter,
and in the presence of the Lorentz Chern-Simons Form, using the so
called Beta Function Favored Constraints \cite{vash}.\footnote{In
earlier works, this made the determination of a D=10 globally
supersymmetric and Lorentz covariant higher derivative Yang-Mills
action possible, to order $\gamma^3$ (see e.g. \cite{more}), an
important result for topologically nontrivial gauge vector field
configurations, as in the case of compactified string theories on
manifolds with topologically nontrivial properties.} This approach
has been detailed to first order in \cite{1}, and to second order
in \cite{3}, so we will not recount it here. We show
that all results fall neatly into place in a very elegant way,
therefore further vindicating the whole original scenario. We note
here that it appears also to work consistently at third order, as
we have proceeded to that order, and that is for yet another work.

\section{Review of Solution and Notation}

The Bianchi identities in Superspace are as follows:

\bea[[\nabla_{[A},\nabla_{B}\},\nabla_{C)}\}=0\eea \

Here we have switched off Yang-Mills fields and the commutator is
given by

 \bea
[\nabla_{A},\nabla_{B}\}=T_{AB}{}^{C}\nabla_{C}+\frac{1}{2}R_{ABd}{}^{e}M_{e}{}^{d}
\eea

A solution must be found in such a way that all if the identities
are simultaneously satisfied. A small alteration in one sector will
change the whole picture. Most of the resulting identities are
listed in \cite{1} and \cite{4}, so we will not list them here. The
second order solution given in part in \cite{3} and \cite{4} to some
extent was based upon an Ansatz for the so called X tensor, as well
as extensive algebraic manipulations. The necessity for introducing
the X tensor was predicted by Gates et. al., \cite{1}. In \cite{3},
and \cite{4}, the following Bianchi identity was not properly
solved:

 \bea
T_{(\a\b|}{}^{\la}R_{|\g)\la}{}_{de}~-~T_{(\a\b|}{}^{g}R_{|\g)}{}_{gde}~
-~\nabla_{(\a|}R_{\b\g)}{}_{de}~=~0\eea

It is crucial to show that
all of the second order torsions and curvatures satisfy this
identity. Also $R^{(2)}{}_{\g gde}$ is required, in order to
complete the set.
 Various ideas, such as finding a new X tensor, imposing constraints on the spinor
derivative $\nabla_{\a}\chi_{\b}$ at second order or adjusting the
super current $A_{abc}$ at second order were considered. We have
found that including these adjustments and constraints is
unnecessary, and might in fact be wrong.

In this paper we find a complete and consistent solution. We also
point out that equation (58) in reference \cite{3} (or equation
(115) in reference \cite{4}) is wrong.

In order to avoid a proliferation of terms we maintain the same
notation and conventions as in \cite{1}, but to avoid
recasting the first order results, we denote all quantities by the
order in the slope parameter

\bea
R_{ABde}=R^{(0)}{}_{ABde}+R^{(1)}{}_{ABde}+R^{(2)}{}_{ABde}+...\nn\eea
\newline
\bea
T_{AD}{}^{G}=T^{(0)}{}_{AD}{}^{G}+T^{(1)}{}_{AD}{}^{G}+T^{(2)}{}_{AD}{}^{G}...\nn\eea

In this work we make some improvements to the notation of
references \cite{3}. For example an apparently
fundamental object is the following:

\bea
\Omega^{(1)}{}_{gef}= L^{(1)}{}_{gef}-\frac{1}{4}A^{(1)}{}_{gef}
\eea
and its spinor derivative which we denote simply  as

\bea \Omega^{(1)}{}_{\a gef} = \nabla_{\g}\{L^{(1)}{}_{gef}-\frac{1}{4}A^{(1)}{}_{gef}\} \eea

We leave it like this for brevity of notation.
The numerical superscript refers to the order of the quantity. A
crucial input at first order is that for the super-current
$A^{(1)}{}_{gef}$.  The choice made for on-shell conditions in
\cite{1} and hence also \cite{3}, is as follows:

\bea A^{(1)}{}_{gef}=i\g \s_{gef \epsilon \tau}T^{mn
\epsilon}T_{mn}{}^{\tau} \eea

In \cite{3},  we proposed the form of the X tensor to
read \bea T^{(2)}{}_{\a\b}{}^{d}~=\s^{pqref}{}_{\a\b}
X_{pqrefd}=-\frac{i\g}{6}\s^{pqref}{}_{\a\b}
H^{(0)}{}^{d}{}_{ef}A^{(1)}{}_{pqr}.\eea

A fundamental result which was used in every Bianchi identity and
which is very lengthy to derive is the following:

\bea
~T^{(0)}{}_{(\a\b|}{}^{\la}\s^{pqref}{}_{|\g)\la}A^{(1)}{}_{pqr}H^{(0)}{}_{def}-
\s^{pqref}{}_{(\a\b|}H^{(0)}{}_{def}\nabla_{|\g)}A^{(1)}{}_{pqr}\nn\\
=-24 \s^{g}{}_{(\a\b|}H^{(0)}{}_{d}{}^{ef}[\Omega^{(1)}{}_{\g
gef}]\eea

We note however in this paper that this result can be achieved
indirectly by using the first order results found in \cite{1}, in
conjunction with the Bianchi identity (3), listed in this paper. We found that the following
dimension one half torsion is given uniquely by:

\bea
T^{(2)}{}_{\a\b}{}^{\la}~=~-~\frac{i\g}{12}\s^{pqref}{}_{\a\b}A^{(1)}{}_{pqr}T_{ef}{}^{\la}.
\eea It was then shown that together with the proposed X tensor
Ansatz as well as equation (8) and other observations and results,
the H sector Bianchi identities as listed in \cite{1}, \cite{2}
could be solved simultaneously with the torsions (10) and (11) as listed below

\bea
T_{(\a\b|}{}^{\la}T_{|\gamma)\la}{}^{d}~-~T_{(\a\b|}{}^{g}T_{|\gamma)g}{}^{d}~
-~\nabla_{(\a|}T_{\b\g)}{}^{d}~=~0 \eea  and  \bea
T_{(\a\b|}{}^{\la}T_{|\g)\la}{}^{\d}~-~T_{(\a\b|}{}^{g}T_{|\g)g}{}^{\d}
~-~\nabla_{(\a|}T_{|\b\g)}{}^{\d} -~\frac{1}{4}R_{(\a\b|
de}\s^{de}{}_{|\g)}{}^{\d} ~=~0 .\eea

We find the second order solutions to (10) to be given by (7) and
the following

\bea \s^{g}{}_{(\a\b|}T^{(2)}{}_{|\g)gd}
=4\g\s^{g}{}_{(\a\b|}\Omega^{(1)}{}_{|\g) gef}H^{(0)}{}_{d}{}^{ef}-
\frac{i\g}{6}\s^{g}{}_{(\a\b|}\s^{pqre}{}_{g|\g)\f}A^{(1)}{}_{pqr}T^{(0)}{}_{de}{}^{\f},
\eea

\bea T^{(2)}{}_{\g ab}=
+2\g[\Omega^{(1)}{}_{\g[a|ef}]H^{(0)}{}_{|b]}{}^{ef}
+\s_{ab~\g}{}^{\f}[\frac{\g}{3}\Omega^{(1)}{}_{\f
gef}H^{(0)}{}^{gef}]\nn\\
-\frac{\g}{6}\s_{[a|}{}^{g}{}_{\g}{}^{\f}\{\Omega^{(1)}{}_{\f
|b]ef}H^{(0)}{}_{g}{}^{ef}
+\Omega^{(1)}{}_{\f gef}H^{(0)}{}_{|b]}{}^{ef}\}\nn\\
-\frac{i\g}{12}A^{(1)}{}_{pqr}\s^{pqrg}{}_{[a|}{}_{\f\la}T^{(0)}{}_{|b]g}{}^{\la}\nn\\
-\frac{i\g}{72}\s_{ab~\g}{}^{\f}\s^{pqreg}{}_{\f\la}A^{(1)}{}_{pqr}T^{(0)}{}_{eg}{}^{\la}\nn\\
\frac{i\g}{144}A^{(1)}{}_{pqr}\s_{[a|}{}^{g}{}_{\g}{}^{\f}[\s^{pqre}{}_{|b]}{}_{\f\la}T^{(0)}{}_{eg}{}^{\la
}+\s^{pqre}{}_{g}{}_{\f\la}T^{(0)}{}_{e|b]}{}^{\la}]\nn\\
\eea

In equation (11), we notice the occurrence of the term

\bea -\nabla_{(\a|}T^{(0)}{}_{|\b\g)}{}^{\d}{}^{(Order 2)}=
[2\d_{(\a|}{}^{\d}\d_{|\b)}{}^{\la}+\s^{g}{}_{(\a\b|}\s_{g}{}^{\d
\la}]\nabla_{|\g)}\chi_{\la}{}^{(2)}.\eea

This was not properly considered in references \cite{3}. In this work we find that there is no need to modify the
spinor derivative of $\chi_{\a}$ at second order so that an
additional constraint on this derivative is unnecessary. For the
solution of (11) we extract after some algebra, and neat
cancelations, the candidates

\bea T^{(2)}{}_{\g g}{}^{\d} =2\g~T^{(0)}{}^{ef}{}^{\d}
\Omega^{(1)}{}_{\g gef} \eea

And \bea R^{(2)}{}_{ \a\b de}
 =-~\frac{i\g}{12}\s^{pqref}{}_{\a\b}A^{(1)}{}_{pqr}R^{(0)}{}_{ef de}.\eea

We now must show that all of the above found results satisfy (3).

\section{New Solution for $R^{(2)}{}_{\la gde}$ }
We must show that we can close equation (3) using the results (7),
(9), (15), and (16). As mentioned, various approaches such as implementing the
previously suggested constraints did not work, nor was there any way
to manipulate the terms using the sigma matrix algebra. Eventually
the following procedure provides a solution. At second order the
Bianchi identity (3) becomes

\bea T^{(0)}{}_{(\a\b|}{}^{\la}R^{(2)}{}_{|\g)\la}{}_{de}+
T^{(2)}{}_{(\a\b|}{}^{\la}R^{(0)}{}_{|\g)\la}{}_{de}~
-T^{(0)}{}_{(\a\b|}{}^{g}R^{(2)}{}_{|\g) g}{}_{de}~-
~T^{(2)}{}_{(\a\b|}{}^{g}R^{(0)}{}_{|\g) g}{}_{de}\nn\\
-~\nabla_{(\a|}[R^{(0)}{}_{|\b\g)}{}_{de}^{Order
(2)}+~R^{(1)}{}_{|\b\g)}{}_{de}^{Order
(2)}+~R^{(2)}{}_{|\b\g)}{}_{de}^{Order (2)}]~=~0.\nn\\\eea

 Using the results listed above we arrive at

 \bea -i\s^{g}{}_{(\a\b|}R^{(2)}{}_{|\g)}{}_{gde}
 +T^{(0)}{}_{(\a\b|}{}^{\la}[
 -~\frac{i\g}{12}\s^{pqrab}{}_{|\g)\la}A^{(1)}{}_{pqr}R^{(0)}{}_{abde}]\nn\\
 -~\frac{i\g}{12}\s^{pqrab}{}_{(\a\b|}A^{(1)}{}_{pqr}T_{ab}{}^{\la}R^{(0)}{}_{|\g)\la}{}_{de}
~+~\frac{i\g}{6}\s^{pqrab}{}_{(\a\b|}
H^{(0)}{}^{g}{}_{ab}A^{(1)}{}_{pqr}R^{(0)}{}_{|\g)}{}_{gde}\nn\\
-\nabla_{(\g|}\{-2i\s^{g}{}_{|\a\b)}\Pi^{(0)+(1)}{}_{gde}+
\frac{i}{24}\s^{pqr}{}_{de}{}_{|\a\b)}A^{(1)}{}_{pqr}\nn\\
-~\frac{i}{12}\s^{pqrab}{}_{|\a\b)}A^{(1)}{}_{pqr}R^{(0)}{}_{abde}\}=0.
\eea

Here we encounter second order contributions from zero order
terms but in solvable form. (That is where we can extract a quantity symmetrized with a sigma matrix)We define

 \bea
 \Pi_{g}{}^{ef}=L{}_{g}{}^{ef}
-~\frac{1}{8}A_{g}{}^{ef} .\eea

Now again using out key relation (8)  we obtain

\bea -i\s^{g}{}_{(\a\b|}R^{(2)}{}_{|\g)}{}_{gde}+2i\g
\s^{g}{}_{(\a\b|}R^{(0)}{}_{abde}[\Omega^{(1)}{}_{|\g) gab}]
-\nabla_{(\g|}\{-2i\s^{g}{}_{|\a\b)}\Pi^{(0)+(1)}{}_{gde}\}\nn\\
-~\frac{i\g}{12}\s^{pqrab}{}_{(\a\b|}A^{(1)}{}_{pqr}T_{ab}{}^{\la}R^{(0)}{}_{|\g)\la}{}_{de}
+~\frac{i\g}{6}\s^{pqrab}{}_{(\a\b|}H^{(0)}{}^{g}{}_{ab}A^{(1)}{}_{pqr}R^{(0)}{}_{|\g)}{}_{gde}\nn\\
+~\frac{i\g}{12}\s^{pqrab}{}_{(\a\b|}A^{(1)}{}_{pqr}[\nabla_{|\g)}R^{(0)}{}_{abde}]~
-~\frac{i}{24}\s^{pqr}{}_{de}{}_{(\a\b|}[\nabla_{|\g)}A^{(1)}{}_{pqr}^{(Order
(2)}]~=~0\nn\\
\eea

Of particular concern and interest is the last term in (20). It was thought that a possible
modification of $A^{(1)}{}_{pqr}$, or a contribution from
$A^{(2)}{}_{pqr}$ would be necessary. Here we may avoid such a modification. In advance we anticipate that the solution
will be as follows:

\bea +i\s^{g}{}_{(\a\b|}R^{(2)}{}_{|\g)}{}_{gde}= +2i\g
\s^{g}{}_{(\a\b|}R^{(0)}{}_{abde}[\Omega^{(1)}{}_{|\g) g}{}^{ab}]
+\nabla_{(\g|}\{2i\s^{g}{}_{|\a\b)}\Pi^{(0)+(1)}{}_{gde}\}{}^{Order (2)}\nn\\
\eea And

 \bea
-~\frac{i\g}{12}\s^{pqrab}{}_{(\a\b|}A^{(1)}{}_{pqr}T_{ab}{}^{\la}R^{(0)}{}_{|\g)\la}{}_{de}
+~\frac{i\g}{6}\s^{pqrab}{}_{(\a\b|}H^{(0)}{}^{g}{}_{ab}A^{(1)}{}_{pqr}R^{(0)}{}_{|\g)}{}_{gde}\nn\\
+~\frac{i\g}{12}\s^{pqrab}{}_{(\a\b|}A^{(1)}{}_{pqr}[\nabla_{|\g)}R^{(0)}{}_{abde}]~
-~\frac{i}{24}\s^{pqr}{}_{de}{}_{(\a\b|}[\nabla_{|\g)}A^{(1)}{}_{pqr}^{(Order
(2)}]=0 \eea

We need to show that (22) does in fact vanish. We must begin with
the Bianchi identity that gives the spinor derivative of
$T_{kl}{}^{\tau}$.

\bea
\nabla_{\g}T_{kl}{}^{\tau}=T_{\g[k|}{}^{\la}T_{\la|l]}{}^{\tau}+
T_{\g[k}{}^{g}T_{g|l]}{}^{\tau}+T_{kl}{}^{\la}T_{\la
\g}{}^{\tau}+T_{kl}{}^{g}T_{g\g}{}^{\tau}-\nabla_{[k|}T_{|l]\g}{}^{\tau}
-R_{kl\g}{}^{\tau}.\eea
 At first order this simplifies to

 \bea \nabla_{\g}T_{kl}{}^{\tau}{}^{Order (1)}=-R^{(1)}{}_{kl\g}{}^{\tau}-\frac{1}{48}
[2H^{(0)}{}_{kl}{}_{g}\s^{g}{}_{\g
\la}\s^{pqr\la\tau}A^{(1)}{}_{pqr}
-\s_{[k|\g\la}\s^{pqr\la\tau}{}(\nabla_{|l]}A^{(1)}{}_{pqr})].\eea

We now write the last term in (22), using the ten dimensional metric
so that the unsolved part becomes

 \bea
-\frac{i}{12}\s^{pqrab}{}_{(\a\b|}\{\g
A^{(1)}{}_{pqr}[T_{ab}{}^{\la}R^{(0)}{}_{|\g)\la}{}_{de}
+T^{(0)}_{ab}{}^{g}R^{(0)}{}_{|\g)}{}_{gde}-\nabla_{|\g)}R^{(0)}{}_{abde}]~\nn\\
+\frac{1}{2}\eta_{ad}~\eta_{be}\nabla_{|\g)}A^{(1)}{}_{pqr}^{Order
(2)}\}=0 \eea

Using the definition of $A^{(1)}{}_{pqr}$ (6), yields
 \bea
+\frac{\g}{12}\s^{pqrab}{}_{(\a\b|}\s_
{pqr\epsilon\tau}T^{kl\epsilon} \{\g
T_{kl}{}^{\tau}[T_{ab}{}^{\la}R^{(0)}{}_{|\g)\la}{}_{de}
+T^{(0)}_{ab}{}^{g}R^{(0)}{}_{|\g)}{}_{gde}-\nabla_{|\g)}R^{(0)}{}_{abde}]~\nn\\
+\eta_{ad}\eta_{be}\nabla_{|\g)}T_{kl}{}^{\tau}\}=0 \eea

We now use equation (24) and the properties of the sigma matrices.
After some algebra we choose to impose a condition on
$R^{(1)}{}_{kl\g}{}^{\tau}$. We require

\bea
R^{(1)}{}_{kl\g}{}^{\tau}=+\frac{\g}{100}T{}_{kl}{}^{\tau}[T_{mn}{}^{\la}R^{(0)}{}_{\g \la}{}^{mn}
+T^{(0)}{}_{mn}{}^{g}R^{(0)}{}_{\g}{}_{g}{}^{mn}-\nabla_{\g}R^{(0)}{}_{mn}{}^{mn}]\nn\\
-\frac{1}{48}[2 H^{(0)}{}_{klg}\s^{g}{}_{\g\la}\s^{rst \la \tau}A^{(1)}{}_{rst}
-2\s_{[k|\g \la}\s^{rst}{}^{\la\tau}\nabla_{|l]}A^{(1)}{}_{rst}] \eea

This can now be added to the list of first order results quoted in \cite{1}.  $R^{(1)}{}_{kl\g}{}^{\tau}$ was not defined in \cite{1}.
We obtain as we required,

\bea R^{(2)}{}_{\g}{}_{gde}= 2\g R^{(0)}{}_{abde}[\Omega^{(1)}{}_{\g
g}{}^{ab}]
+2\nabla_{\g}\{\Pi^{(0)+(1)}{}_{gde}\}{}^{Order (2)}\nn\\
\eea

As a check we can also examine another Bianchi identity. The
following Bianchi identity  also includes $R^{(2)}{}_{\a bde}$:

\bea
 \frac{1}{4}R_{(\a|}{}_{amn}\s^{mn}{}_{|\b)}{}^{\g}
+T_{\a\b}{}^{g}T_{ga}{}^{\g}+T_{\a\b}{}^{\la}T_{\la
 a}{}^{\g}+
 T_{a(\a|}{}^{\la}T_{|\b)\la}{}^{\g}-T_{a (\a|}{}^{g}T_{|\b)
 g}{}^{\g}\nn\\-\nabla_{(\a|}T_{|\b)a}{}^{\g} -\nabla_{a}T_{\a
 \b}{}^{\g}=0
\eea

This Bianchi identity after some cancelations results in the
following expression:

\bea \frac{1}{4}R^{(2)}{}_{(\a|amn}\s^{mn}{}_{|\b)}{}^{\g}+
2\g\{\nabla_{(\a|}\Omega^{(1)}{}_{aef}\}[-\frac{1}{4}R^{(0)}{}_{ef}{}^{mn}\s_{mn
|\b)}^{\g}]\nn\\ +i\s^{g}{}_{\a\b}T^{(2)}{}_{g a}{}^{\g}-
\frac{i\g}{6}\s^{pqref}{}_{\a\b}A^{(1)}{}_{pqr}H^{(0)g}{}_{ef}T^{(0)}{}_{ga}{}^{\g}
+T^{(0)}{}_{\a\b}{}^{\la}T^{(2)}{}_{\la a}{}^{\g}\nn\\
-2\g
T^{(0)}{}_{ef}{}^{\g}[\nabla_{(\a|}\nabla_{|\b)}\{\Omega^{(1)}{}_{aef}\}]
\nn\\
+\frac{i\g}{12}\s^{pqref}{}_{\a\b}\{\nabla_{a}A^{(1)}{}_{pqr}\}T^{(0)}{}_{ef}{}^{\g}
+\frac{i\g}{12}\s^{pqref}{}_{\a\b}A^{(1)}{}_{pqr}[\nabla_{a}T^{(0)}{}_{ef}{}^{\g}]\nn\\
+[\d_{(\a|}{}^{\la}
\d_{|\b)}{}^{\f}+\s^{g}{}_{\a\b}\s_{g}{}^{\la\f}]\nabla_{a}\chi_{\f}{}^{Order
2} \nn\\+\frac{1}{4}\s^{nm}{}_{(\a|}{}^{\g}\nabla_{|\b)}\Pi_{amn}{}^{Order (2)} =~0\nn\\\eea

This identity also predicts the same form for $R^{(2)}{}_{\a amn}$.
However it also includes a great deal of other information which
we plan to include in another letter.

\section{Conclusions}

We have found a consistent and manifestly supersymmetric solution to
the Bianchi identities for D=10, N=1 supergravity, with string
corrections to second order in the slope parameter. We have
reaffirmed the results and the proposed X tensor of  \cite{3} and
\cite{4}, and we have solved the remaining previously intractable
curvature. We have used the first order result for
$A^{(1))}{}_{pqr}$ as first given in \cite{1}. We have not modified
it at second order. We have not imposed the constraint

 \bea T_{\a b}{}^{\d}=-\frac{1}{48}\s_{b\a \la}\s^{pqr
\la\d}A_{pqr}\eea

at second order. This is a conventional
constraint and so could have been be imposed to all orders. However
we dropped it in favor of requiring an adjustment to $T_{\a
b}{}^{\d}{}^{(2)}$ as given by equation (15).

\section{Acknowledgement}
We would like to thank S. J. Gates, Jr. for useful discussions.
The work of S.B. has been supported in part by the European
Community Human Potential Program under contract
MRTN-CT-2004-005104 \textit{``Constituents, fundamental forces and
symmetries of the universe''}.

\section{Appendix I}

Here for convenience we list the torsions curvatures and H sector
results to second order, simply by including the results found at
first order in \cite{1}.

\bea H_{\a\b\g}=0 + Order(\g^{3})\eea

\bea H_{\a\b d}=+\frac{i}{2}\s_{d \a\b}
+4i\g\s^{g}{}_{\a\b}H_{\g}{}^{ef}H_{d}{}^{ef}\nn\\
~\s_{\a\b}{}^{g}[{8i\g}H^{(0)}{}_{def}L^{(1)}{}_{g}{}^{ef}~ -~i\g
H^{(0)}{}_{def}A^{(1)}{}_{g}{}^{ef}]\nn\\~+~
\s^{pqref}{}_{\a\b}[\frac{i\g}{12}H^{(0)}{}_{def}A^{(1)}{}_{pqr}]+Order(\g^{3})
\eea

\bea H_{\a ab}=+2i\g[-\s_{[a|}{}_{\a\b}T_{ef}{}^{\b}G_{|b]}{}^{ef}-
2\s_{e}{}_{\a\b}T_{f[a|}{}^{\b}G_{|b]}{}^{ef}]\nn\\
~2\g[\nabla_{\a}(H^{(0)}_{[a|ef}H^{(0)}{}_{|b]}{}^{ef}~
-~\s_{ab\a}{}^{\f}\nabla_{\f}(H^{(0)}{}_{gef}H^{gef})]\nn\\
~+2i\g\s_{[a|\a\f}T_{ef}{}^{\f}\Pi^{(1)}{}_{|b]}{}^{ef}
-~2i\g\s_{ab\a}{}^{\la}\s_{g\la\f}T_{ef}{}^{\f}\Pi^{(1)gef}\nn\\
-~\frac{\g}{6}\s^{g}{}_{[a|\a}{}^{\f}\s_{|b]\la\f}T_{ef}{}^{\la}\Pi^{(1)}{}_{g}{}^{ef}
-~\frac{\g}{6}\s^{g}{}_{[a|\a}{}^{\f}\s_{g\la\f}T_{ef}{}^{\la}\Pi^{(1)}{}_{|b]}{}^{ef}\nn\\
~-~4\g R^{(1)}{}_{\a[a|}{}^{ef}H^{(0)}{}_{|b]ef}+T^{(2)}{}_{\a
ab}+Order(\g^{3})\eea

\bea T_{\a\b}{}^{g}=
i\s_{\a\b}{}^{g}-\frac{i\g}{6}\s^{pqref}{}_{\a\b}
H^{(0)}{}^{d}{}_{ef}A^{(1)}{}_{pqr}+Order(\g^{3})\eea

\bea T_{abc}=-2L_{abc}\eea

\bea
T_{\a\b}{}^{\g}=-[\d_{(\a|}{}^{\g}\d_{|\b)}{}^{\d}+\s^{g}{}_{\a\b}\s_{g}{}^{\g
\d}]\chi_{\d}~-~\frac{i\g}{12}\s^{pqref}{}_{\a\b}A^{(1)}{}_{pqr}T_{ef}{}^{\g}+Order(\g^{3})
\eea

\bea T_{\a g}{}^{\d} =
 -\frac{1}{48}\s_{g \a \f}\s^{pqr}{}^{\f\d}A^{(1)}_{pqr}
 +2\g~T^{(0)}{}^{ef}{}^{\d}
\Omega^{(1)}{}_{\a gef} +Order(\g^{3})\eea

\bea \s^{g}{}_{(\a\b|}T^{(2)}{}_{|\g)gd}
=4\g\s^{g}{}_{(\a\b|}\Omega_{|\g) gef}H^{(0)}{}_{d}{}^{ef}-
\frac{i\g}{6}\s^{g}{}_{(\a\b|}\s^{pqre}{}_{g|\g)\f}A^{(1)}{}_{pqr}T^{(0)}{}_{de}{}^{\f}
\eea
 Or symmetrized,

\bea T{}_{\g ab}=
+2\g[\Omega^{(1)}{}_{\g[a|ef}]H^{(0)}{}_{|b]}{}^{ef}
+\s_{ab~\g}{}^{\f}[\frac{\g}{3}\Omega^{(1)}{}_{\f
gef}H^{(0)}{}^{gef}]\nn\\
-\frac{\g}{6}\s_{[a|}{}^{g}{}_{\g}{}^{\f}\{\Omega^{(1)}{}_{\f
|b]ef}H^{(0)}{}_{g}{}^{ef}
+\Omega^{(1)}{}_{\f gef}H^{(0)}{}_{|b]}{}^{ef}\}\nn\\
-\frac{i\g}{12}A^{(1)}{}_{pqr}\s^{pqrg}{}_{[a|}{}_{\f\la}T^{(0)}{}_{|b]g}{}^{\la}\nn\\
-\frac{i\g}{72}\s_{ab~\g}{}^{\f}\s^{pqreg}{}_{\f\la}A^{(1)}{}_{pqr}T^{(0)}{}_{eg}{}^{\la}\nn\\
\frac{i\g}{144}A^{(1)}{}_{pqr}\s_{[a|}{}^{g}{}_{\g}{}^{\f}[\s^{pqre}{}_{|b]}{}_{\f\la}T^{(0)}{}_{eg}{}^{\la
}+\s^{pqre}{}_{g}{}_{\f\la}T^{(0)}{}_{e|b]}{}^{\la}]+Order(\g^{3})\nn\\
\eea

\bea R_{\a\b de} =
-2i\s^{g}{}_{\a\b}\Pi_{gde}{}^{(1)}+\frac{i}{24}\s^{pqref}{}_{\a\b}A_{pqr}{}^{(1)}\nn\\
-~\frac{i\g}{12}\s^{pqref}{}_{\a\b}A^{(1)}{}_{pqr}R_{ef
de}+Order(\g^{3})
 \eea
 Where

 \bea
 \Pi^{(1)}{}_{g}{}^{ef}=L^{(1)}{}_{g}{}^{ef}
-~\frac{1}{8}A^{(1)}{}_{g}{}^{ef} \eea

\bea
 R_{\a gde}= -i\s_{[d|}{}_{\a \f}T_{g |e]}{}^{f} +i\g\s_{[g|}{}_{\a~
 \f}T_{kl}{}^{\f}R^{kl}{}_{|de]}\nn\\
 + 2\g
 R^{(0)}{}_{abde}[\Omega^{(1)}{}_{\a
g}{}^{ab}]
+2\nabla_{\a}\{\Pi^{(0)+(1)}{}_{gde}\}{}^{Order (2)}+Order(\g^{3})\nn\\
\eea

\bea A_{abc}=i\g \s_{gef \g \la}T^{mn \g}T_{mn}{}^{\la}
\eea

\bea
R^{(1)}{}_{kl\g}{}^{\tau}=+\frac{\g}{100}T{}_{kl}{}^{\tau}[T_{mn}{}^{\la}R^{(0)}{}_{\g \la}{}^{mn}
+T^{(0)}{}_{mn}{}^{g}R^{(0)}{}_{\g}{}_{g}{}^{mn}-\nabla_{\g}R^{(0)}{}_{mn}{}^{mn}]\nn\\
-\frac{1}{48}[2 H^{(0)}{}_{klg}\s^{g}{}_{\g\la}\s^{rst \la \tau}A^{(1)}{}_{rst}
-2\s_{[k|\g \la}\s^{rst}{}^{\la\tau}\nabla_{|l]}A^{(1)}{}_{rst}] \eea

The spinor derivative of $L_{abc}$ is solved and available from a
Bianchi identity. We will list it in a later paper. $R^{(2)}{}_{kl\g}{}^{\tau}$ if it exits will likely show up from
third order calculations of (3).


\begin{thebibliography}{99}


\bibitem{1} S. Bellucci, D.A. Depireaux and S.J. Gates, Jr., Phys. Lett.
B238 (1990) 315 ;  S.J. Gates, Jr., A. Kiss, W. Merrell, JHEP 0412
(2004) 047.

\bibitem{2} S.J. Gates, Jr. and H. Nishino, Nucl. Phys. B291 (1987) 205; ibid.
  Phys. Lett. B173 (1986) 52  ;  S.J. Gates, Jr. and S. Vashakidze, Nucl. Phys. B291 (1987)
  172; S. Bellucci and S.J. Gates, Jr., Phys. Lett. B208 (1988) 456.



\bibitem{3} S. Bellucci and D. O'Reilly, Phys. Rev. D 73, 065009
(2006); hep-th/0603033

\bibitem{4} D. O'Reilly, PhD. Thesis, The Graduate Center, City
University of New York, December 2005;  hep-th/0601184


\bibitem{vash} S. Bellucci, S.J. Gates, Jr., B. Radak and S. Vashakidze, Mod. Phys. Lett. A4
(1989) 1985; see also S. Bellucci, R.N. Oerter, Nucl. Phys. B 363
(1991) 573; S. Bellucci, Mod. Phys. Lett. A5 (1990) 2253; ibid.
Nucl. Phys. B313 (1989) 220; ibid. Phys. Lett. B227 (1989) 61;
ibid. Mod. Phys. Lett. A3 (1988) 1775; ibid. Prog. Theor. Phys. 79
(1988) 1288; ibid. Z. Phys. C36 (1987) 229; ibid. Z. Phys. C41
(1989) 631.


\bibitem{more} S. Bellucci, Z. Phys. C50 (1991) 237; ibid. Prog. Theor. Phys. 84 (1990) 728.





\end{thebibliography}
\end{document}